\documentstyle[preprint,aps]{revtex}

\begin{document}
\title{Type I intermittency in a dynamical system with dichotomous parameter change}
\author{J.J.\.Zebrowski and $^{+}$R.Baranowski}
\address{Faculty of Physics, Warsaw University of Technology, Warszawa, Poland;\\
$^{+}$National Institute of Cardiology at Anin}
\date{\today}
\maketitle
\pacs{05.45.-a, 87. , \ 82.40.Bj, 87.19.Hh}

\begin{abstract}
In type I intermittency, simple models known for at least twenty years show
that a characteristic u-shaped probability distribution is obtained for the
laminar phase length. We have shown elsewhere that, for some cases of
pathology, the laminar phase length distribution characteristic for type I
intermittency may be obtained in human heart rate variability data. The
heart and its regulatory systems are presumed to be both noisy and
nonstationary. Although the effect of additive noise on the laminar phase
distribution in type I intermittency is well known, neither the effect of
multiplicative noise nor of nonstationarity (i.e. changes of the control
parameter with the time) have been studied. In this paper, we first discuss
the properties of two classes of models of type I intermittency: a) the
control parameter of the logistic map is changed dichotomously from a value
within the intermittency range to just below the bifurcation point and back;
b) the control parameter is changed randomly within the same parameter range
as in the model class a). We show that the properties of both models are
importantly different from those obtained for type I intermittency in the
presence additive noise as obtained by Hirsch twenty years ago. The two
models help explain some of the features seen in the intermittency in human
heart rate variability.
\end{abstract}

\bigskip

Intermittency in stationary dynamical systems is a broad area of research of
constant interest (over 1000 papers published so far). \ The phenomenon
appears whenever a dynamical system is close to one of several common types
of bifurcation. The simplest kind of intermittency was classified by Pomeau
and Manneville \cite{Ott} as type I, II and III and shown to be due to the
proximity of the system to the saddle-node, the Hopf and the reverse period
doubling bifurcation, respectively. In these basic types of intermittency,
the evolution with the time of the system may be divided into ranges of the
time in which the behavior of the system is (almost) regular - the laminar
phases - and ranges in which chaotic bursts occur. Other, less common types
of regular intermittency have also been identified \cite{type X}\cite{type V}%
. Chaos-chaos intermittency is due to crisis phenomena occurring in the
system \cite{Ott} and on-off intermittency is due to a symmetry breaking
bifurcation \cite{on-off}. Thus, identification of the type of the
intermittency observed may yield important information about the system by
specifying the bifurcations possible for its dynamics.

Identification of the type of intermittency in experiments and numerical
analysis is made by studying its statistical properties \cite{Hirsch}\cite
{Kacperski}\cite{BeckerBenner}\cite{Jin-Hang Cho} where usually the
dependence of the average laminar phase length $<l>$ on the value of the
control parameter and the probability distribution of the laminar phase
lengths $P(l)$ as a function of that parameter are studied. When the shape
of the map is known or may be measured, reinjection probability density
(RPD) analysis may be performed \cite{RPD1996}. Note that, in many
experimental situations, the map is unknown and, particularly in biology and
astrophysics, the control parameter is not accessible and so only the
distribution $P(l)$ may be studied.

Another factor important in experimental measurement is noise. Additive
noise is then unavoidable and much attention to the effect of additive noise
has been given on intermittency in dynamical systems \cite{Hirsch}\cite
{BeckerBenner}\cite{Jin-Hang Cho}\cite{Frank}\cite{C.Kim}\cite{Feng}\cite
{chil-min}\cite{chil-mina}\cite{MOKim}. \ The most important feature of the
effect of noise on type I intermittency is that the well known, textbook, \
symmetric U-shape of the $P(l)$ curve is changed: the right peak of the
distribution broadens, becomes smaller and gains a long tail extending
towards larger laminar phase lengths \cite{Hirsch}\cite{BeckerBenner}.
Notably, the distribution as a whole becomes narrower. More recently, \ the
effect of additive noise was studied as the control parameter of a
simplified, generic model of intermittency was changed from pre-bifurcation
values to those above the saddle node bifurcation point and non-trivial
results were obtained \cite{Jin-Hang Cho}.

Recently, we studied \cite{ladek} the effect of multiplicative noise on the
type I\ intermittency which occurs close to the period-3 window in the
logistic map. We showed that small, random changes of the control parameter
within the bounds of intermittency broaden the $P(l)$ distribution instead
of narrowing it as does the additive noise.

We have found \cite{ActaPhysInterm}\cite{ladek} that in some cases of human
heart rate variability type I intermittency may occur. We showed that the
laminar phase lengths - defined as parts of the 24-hour long heart rate
variability time series with a large standard deviation - have a probability
distribution which is U-shaped albeit strongly asymmetric and with a long
tail extending towards large laminar phase lengths. In many cases to the
right of the main U-shape distribution there occurs a secondary peak in the
distribution. Some of these distortions may be partly due to the shape of
the underlying, unknown mapping involved but certainly may not be explained
by the effect of additive noise often expected in living systems. Judging by
the clinical data of the cases we have examined, it appears that the
occurrence of intermittency in human heart rate variability is related to
supraventricular or ventricular arrhythmia. This holds also when the number
of ectopic beats is low enough so that from the medical point of view the
recording may be qualified as normal. Arrhythmia may shift the state of the
system (i.e. change its control parameters) for a certain time. To
understand what effect such a change of parameters may have on a system in
type I intermittency, we propose simple models based on the logistic map in
which {\it dichotomous parametric change} is active. We show that the
effects of the dichotomous change of the control parameter explain the main
features of our results found in heart rate variability: the long tail
outside the U-shaped main distribution as well as the secondary peaks which
occur occasionally in heart rate variability data. The paper is organized as
follows. Firstly, we briefly review the medical data used in this study. We
next define the models with dichotomous parametric change. We then present
the results obtained for heart rate variability and the results of our
calculations. A brief discussion and conclusions section follows.

\section{The medical data}

\bigskip

Heart rate variability data was extracted from 24-hour Holter device ECG
recordings using the 563 Del\ Mar Avionics system at the National Institute
of Cardiology (Warszawa, Poland). All data was checked by a qualified
cardiologist: normal beats were detected, artifacts were deleted and
arrhythmias were recognized. The data was sampled at 256 Hz.

The patient BLT had a permanent atrial tachycardia that was conducted to the
ventricles with a varying extent of the atrio-ventricular block. No
ventricular ectopy was observed. The patient FTCH had sinus and atrial heart
rhythm with an extremely increased number of ventricular arrhythmia (around
50\% of heart cycles were ectopic). Additionally, the presence of arrhythmia
influnced the sinus or atrial rhythm due to retrograde conduction. The
patient WJCK had sinus rhythm and ectopic beats (7\% of all cycles)
originating in the ventricles and in the atria.

\section{Models of nonstationary intermittency}

The logistic map model $x_{n+1}=ax_{n}(1-x_{n})$ where $x_{n}\in \lbrack 0,1]
$ , $a$ is the control parameter and $n$ the iteration index is well known 
\cite{Ott}. \ This map has been widely used to model the properties of type
I intermittency although this phenomenon will occur in any system in which
the saddle-node bifurcation occurs at a control parameter range just below
bifurcation point. We modified this basic model to generate  nonstationary
intermittency data.

For control parameter values larger than 3.5699456..... (the accumulation
point \cite{Ott}), periodic windows occur due to tangential bifurcations.
These periodic windows are dense for parameter value lying between the
accumulation point and $a=4$. Type I intermittency occurs for control
parameter values slightly below that at which the tangential bifurcation of
the given periodic window occurs. The intermittency occurring close to the
widest periodic window - the period 3 window - was chosen for this study.
The critical value of the control parameter $a$ at which the period 3 orbit
occurs was determined numerically to be $a_{c}=3.8284271271245$. The
difficulty in assessing the critical value of the control parameter was that
as the bifurcation point was approached the length of the laminar phase
(i.e. the number of iterations corresponding to the periodic state between
chaotic bursts) grows to infinity. At $a=a_{c}$ the length of the periodic
orbit obtained exceeded 100 0000 iterations. The critical value of the
control parameter was determined numerically because it was important for
this study that - given the large accuracy of the control parameter used for
the control parameter - the position of the bifurcation point which occurs
for the processor in the computer on which the calculations were performed
be used in the analysis. The exact position of the bifurcation point may be
slightly machine and compiler dependent due to roundoff errors.

The value of the control parameter, for which the maximum of the length of
the laminar phase would attain a manageable value, was sought. After several
trials, the control parameter $a=3.828$ was assumed. At this value of the
control parameter the maximum laminar phase was 44 iterations.

Nonstationarity may be introduced in many ways. Below, the following models
of nonstationary intermittency were considered.

\subsection{Dichotomous parameter change models}

The dichotomous parameter change models were obtained using the standard
formula for the logistic map. However, the value of the control parameter $a$
was cycled with a period of $n+m$ iterations where $n$ was the number of
iterations for which $a$ equaled $3.828$ and $m$ was the number of
iterations for which $a$ equaled either $3.8284271225$ (model d1) or $%
3.828427128$ (model d2). Both $n$ and $m$ were varied from $1$ to $20$. In
some cases the range of $m$ was extended to 100.

The motivation to use these models was the supposition that some of the
results obtained for human heart rate variability may be due to a response
of the regulatory system to premature ventricular beats (instances of
arrhythmia). When such heart beats occur and the regulatory system of the
heart is in the intermittency state (i.e. close to a tangential bifurcation)
then the occurrence of a premature beat may shift the control parameter
closer towards the bifurcation point (model d1) or slightly beyond it (model
d2).

As for the random parametric noise models discussed in ref.\cite{ladek}, the
results obtained here for the dichotomous parameter change were only weakly
dependent on whether the control parameter assumed the value just below the
bifurcation point (model d1) or just above it (model d2). For this reason
only the calculations for model d1 will be discussed here. Note, however,
that all the calculations discussed below were also performed for model d2.

\subsection{Dichotomous parametric noise model}

In this model again the standard formula for the logistic map was used but
the value of the control parameter at every iteration was changed randomly
between two values: $3.828$ and $3.8284271225$. The model had a secondary
control parameter: the probability with which these two values were chosen.

\section{Calculation of the distribution of the laminar phases}

\bigskip

Given a time series either from measurement or obtained through calculations
in a model, the laminar phases must be found and their distribution
calculated. In the generic, simplified model of type I intermittency \cite
{Ott} the subsequent iterations during the laminar phase of intermittency
have a very low standard deviation. This is because the generic model
describes the behavior of the system in the narrow channel close to the
bifurcation point. In an intermittency state occurring close to period-k
window there will be k such channels so that, except for k = 1, the
resultant time series within the laminar phases will have a large standard
deviation.

We calculated the average deviation of each time series studied here within
a sliding window of j data points, j = 2-10 (windows up to 100 data points
were tested)\cite{ActaPhysInterm}\cite{ladek}. A histogram of the resultant
time series of the window average deviation was calculated. Maxima in this
histogram defined the phases i.e. those of the laminar and of chaotic
bursting behavior. The minimum between such maxima was used as the
discriminating criterion and - when the average deviation was larger than
this the coding level - the number of iterations between the crossings of
the coding level defined the length $l$ of the laminar region. Finally, the
probability density distribution $P(l)$ was calculated as a normalized
histogram of $l$.

To use the above described algorithm one needs to define the sliding window
length. Tests carried out with stationary intermittency time series
calculated using the logistic map show that the window length should be
approximately equal to the period of the orbit in the adjacent periodic
window. If the length of the sliding window is too small compared with the
period of the orbit, a Poisson-like distribution is obtained for $P(l)$\
falsely indicating no intermittency. If, on the other hand, the window is
too long then some averaging results and a false tail extending towards long
laminar phase lengths appears. Below, all data was analyzed using windows
either 3 or 5 data points in length.

\section{ Results}

\subsection{Intermittency in human heart rate variability}

\bigskip

The most common type of laminar length distribution found in our studies is
shown in fig. 1 (compare fig.1 in ref.\cite{ladek}) for window size 3 RR
intervals and in fig.2 for the patient WJCK. It was found that in most cases
studied here 3 or 5 data point length were the optimal lengths of the
sliding window. Larger values averaged the time series in such a way that
the details of the time series were lost. It can be seen that an almost
perfect shape of the probability distribution for intermittency type I was
obtained in both cases except that there was a tail to the right of the
characteristic U- shaped distribution. In other examples of such a shape of
the distribution the tail in the distribution extended still further towards
longer laminar lengths.

A double peaked distribution of laminar lengths was obtained for window
lengths 3 and 5 intervals (the latter is depicted in fig.3) for the patient
FTCH. Other examples of such split peak in the $P(l)$ distribution were
found for different patients suffering from different forms of arrhythmia.

The results shown here similarly as those published earlier in \cite
{ActaPhysInterm} and in \cite{ladek} indicate that the distribution of
laminar lengths obtained from the examples of human heart rate variability
are very typical for such distributions obtained for the standard models of
type I intermittency \cite{Ott}\cite{Hirsch} except for the following
features: there is a long tail extending towards long laminar phase lengths
present in all cases and in some of them also a second peak or multiple
peaks are obtained. Using numerical models, two possible reasons for these
additional features are examined below.

\subsection{\protect\bigskip Intermittency in the presence of dichotomous
noise}

\bigskip

Fig.4 depicts the distribution of laminar phases for the logistic map with
random dichotomous parametric noise. It can be seen that when the two values
of the control parameter were chosen with an equal probability (crosshatched
bars) the effect was very similar to that of the random parametric noise of
ref.\cite{ladek}: the distribution of laminar phases becomes broader than
for the stationary case (black bars) and smaller. However, when the
probability of choosing the prebifurcation value $3.8284271225$ was 25 \%
then the $P(l)$ distribution (open bars)  the shape of the stationary
distribution s retained but a tail extending towards long laminar phase
lengths appears.

\subsection{Intermittency in the presence of dichotomous parameter change}

\bigskip

For $n=1$ and $m$ changed from $1$ upwards, it was found that \ for $%
m=2+3k,\;k=0,1,2,3,....$ the period-3 state was obtained although - without
nonstationarity - for $a=3.8284271225$ intermittency was found for at least $%
10^{5}$ iterations. For other values of $m$ the only effect of dichotomous
parameter change was to increase the width of the distribution of the
laminar phase lengths. Examples of such behavior are shown in fig.5 for $%
m=1,6,10.$ Similarly, for $n=2$ with $m=2$ or $\ m=3$, itermittency with a
broadened distribution of \ laminar phase lengths was obtained. However, for 
$\ n=2$ and all other values of $\ m$ the period-3 state was obtained
without intermittency. Also for $n=3$ and $4$, although no periodic states
were found, the only effect of the dichotomous parameter change was to
broaden the distribution for all values of $m$.

A new phenomenon occurred for $n=5.$ For $m<10$, again just the broadening
of the distribution was found - similar to that in fig.5. Fig.6 depicts the
distributions for the stationary case (black bars), $m=10$ (cross hatched
bars) and for $m=25$ (white bars). It can be seen that a large splitting of
the right peak in the distribution occurs. A further increase of $m$
resulted in an increase of the width of the distribution but the right hand
peaks decreased in size and drew closer together. At $m=75$ the maximum
phase length was slightly less than 225 iterations but right peak was very
broad and its splitting was only weakly visible.

For $n=10$ and $m=2$ a strong splitting of the right peak was observed
(Fig.7). It can be seen that the right peak of this distribution for the
nonstationary case is split\ into two maxima in such a way that the left of
these falls somewhat to the left of the maximum laminar phase length for the
stationary case. Surprisingly for $m=3$ no splitting occurred but the
distribution had the same shape as for the stationary case. However, the
maximum laminar phase length was slightly less than for $m=2$ (not shown).
For $m=5,$ a splitting similar to that obtained for $m=2$ occurred with the
distribution broadening up to laminar phase length 60. A further increase of 
$m$ resulted in a decrease of the splitting with a simultaneous broadening
of the distribution and of its right hand peak so that at $m=30$ the
splitting was only barely visible.

\section{Discussion and Conclusions}

\bigskip

Recently, we have demonstrated that 24-hour human heart rate variability may
exhibit the main statistical features of type I intermittency \cite
{ActaPhysInterm}\cite{ladek}. We present here two kinds of data. The first
are three characteristic examples of \ type I\ intermittency found by us in
heart rate variability measurements. The second are the results of
calculations in simple models of type I intermittency in a nonstationary
system. The results obtained form these models seem to be interesting in
themselves: new properties of a system in type I intermittency with
parametric noise and under the action of a dichotomous parameter change were
obtained. However, an important motivation in defining these models was the
attempt to understand the reasons for certain deviations of the properties
of the intermittency found in heart rate variability from the textbook
descriptions \cite{Ott}. Two reasons for these deviations were considered:
the effect of dichotomous parametric noise and the effect of abrupt,
dichotomous control parameter change such as may be due to episodes of
arrhythmia.

The models considered here are based on the well known logistic equation
which has been widely used as the primary model of type I intermittency.
These models are of two kinds: in the first, a random, multiplicative,
dichotomous noise is applied in such a way that the system approaches
randomly very close to the bifurcation point. Such a model would correspond
to arrhythmia occurring erratically in the time. In the second category of
models, the control parameter is dichotomously cycled with the period n+m: n
iterations at a predefined control parameter value inside the intermittency
range and the next m iterations at a control parameter value either very
close the bifurcation point but in intermittency or just inside the periodic
window. It was found that only small quantitative differences were obtained
depending on whether the system goes through the saddle-node bifurcation or
not. Such a class of models would correspond to an arrhythmia occurring in a
regular way, every certain number of normal heartbeats.

Comparison of the properties of the models with the results obtained for
human 24-hour heart rate variability show that a) the long tails in the
laminar length probability density distributions are most probably due to
multiplicative (parametric) noise and not to additive noise, b) the double
or triple peaks in such distributions found in some cases for heart rate
variability are due to the interplay between the number of natural heart
beats and the number of ectopic beats (arrhythmia). Note that, for the
extremely simple models presented here, even very few iterations of the
system with the control parameter in the vicinity of the bifurcation point
interspersed between several iterations performed far away from that point
introduce a strong splitting of the peaks of the laminar phase lengths
distribution. Although no discrete map models of the complete heart rate
variability control system exist, nonlinear dynamics shows that many
features found in simple models such as the logistic map are universal and
are often found in more elaborate models. Note, that the circle map \cite
{Ott} is most probably a more adequate class of maps to describe the
properties heart rhythm \cite{Glass}. However, \ we need to remember that
type I intermittency occurs also in circle maps and that the basic
properties of  type I intermittency are map independent. In choosing the
logistic map as the core of the nonstationary models discussed here we have
avoided the cumbersomeness of two-parameter models.

The results presented here seem to indicate that the heart rate variability
regulating system - at least for the types of pathology studied in the
examples given - may be permanently close to a saddle-node bifurcation.

\section{Acknowledgments}

\bigskip

This paper was supported by KBN grant no 5 P03B 001 21.

\section{Figure captions}

\bigskip

Fig.1 The laminar phase length probability density found for the 24-hour
heart rate variability recording for the patient BLT.

Fig.2 The laminar phase length probability density found for the 24-hour
heart rate variability recording for the patient WJCK.

Fig.3 The laminar phase length probability density found for the 24-hour
heart rate variability recording for the patient FTCH. Note the
characteristic splitting of the righthand peak of the distribution.

Fig.4 Distribution of laminar phases for the logistic map with random
dichotomous parametric noise: black bars 85 \% probability of a=3.828,
crosshatched bars 50 \% probability of a=3.828. White bars - stationary case
without noise.

Fig.5 Distribution of laminar phase lengths obtained for the the dichotomous
parametric change model with $n=1$ and $m=1,6,10$ as marked and for the
sationary case S. The left maxima of all distributions coincide.

Fig.6 Distribution of laminar phase lengths with a split righthand peak
obtained for the the dichotomous parametric change model with $n=5$ and $%
m=10 $ (cross hatched bars) and for $m=25$ (white bars). The black bars
depict the result for the statioanry case. Peaks at right maxima are
labelled with the value of $m$ and the left maxima of all distributions
coincide.

Fig.7 Distribution of laminar phase lengths with a split righthand peak
obtained for the the dichotomous parametric change model with $n=10$ and $%
m=2,5$ and $25$ (cross hatched, white and horizontally hatched bars,
respectively). Stationary case - black bars.

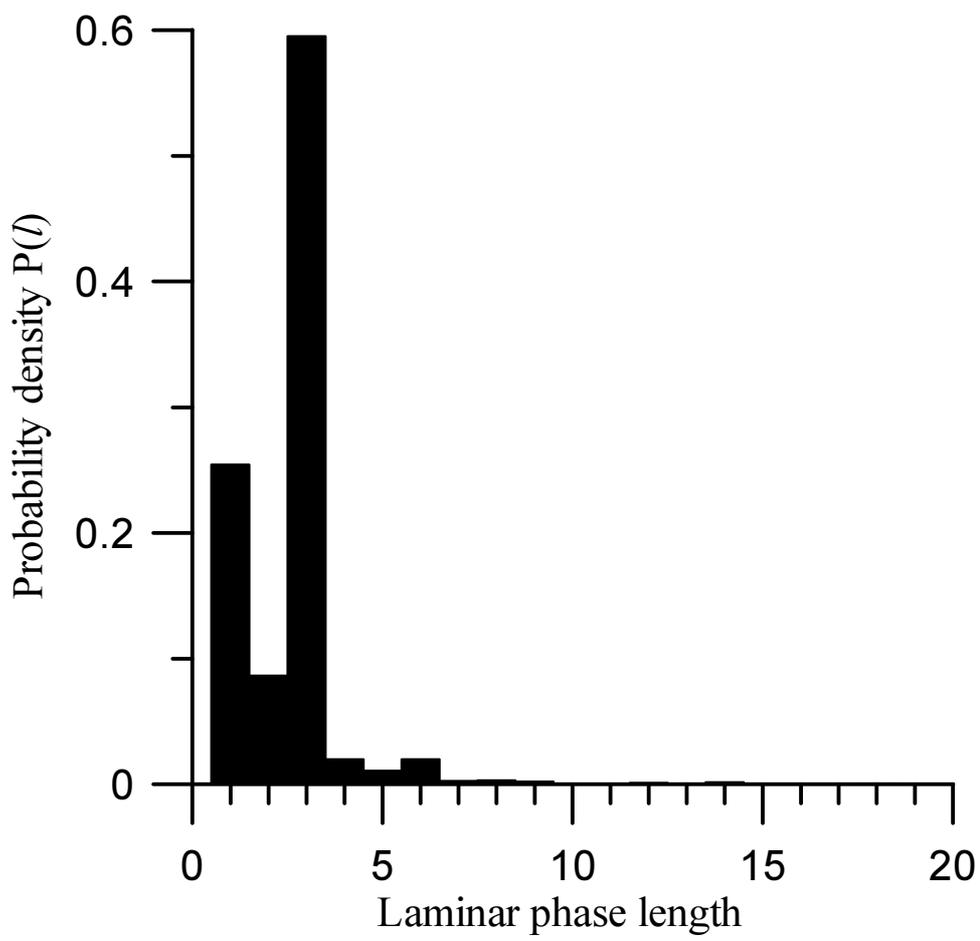
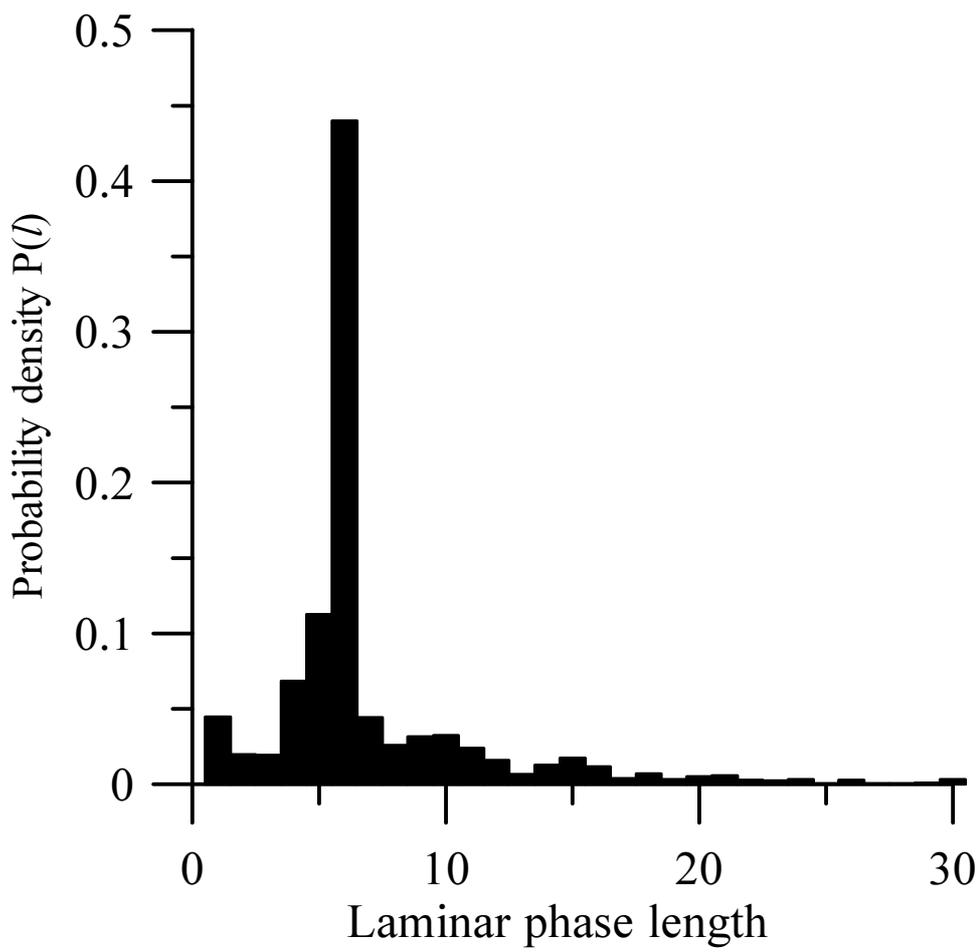
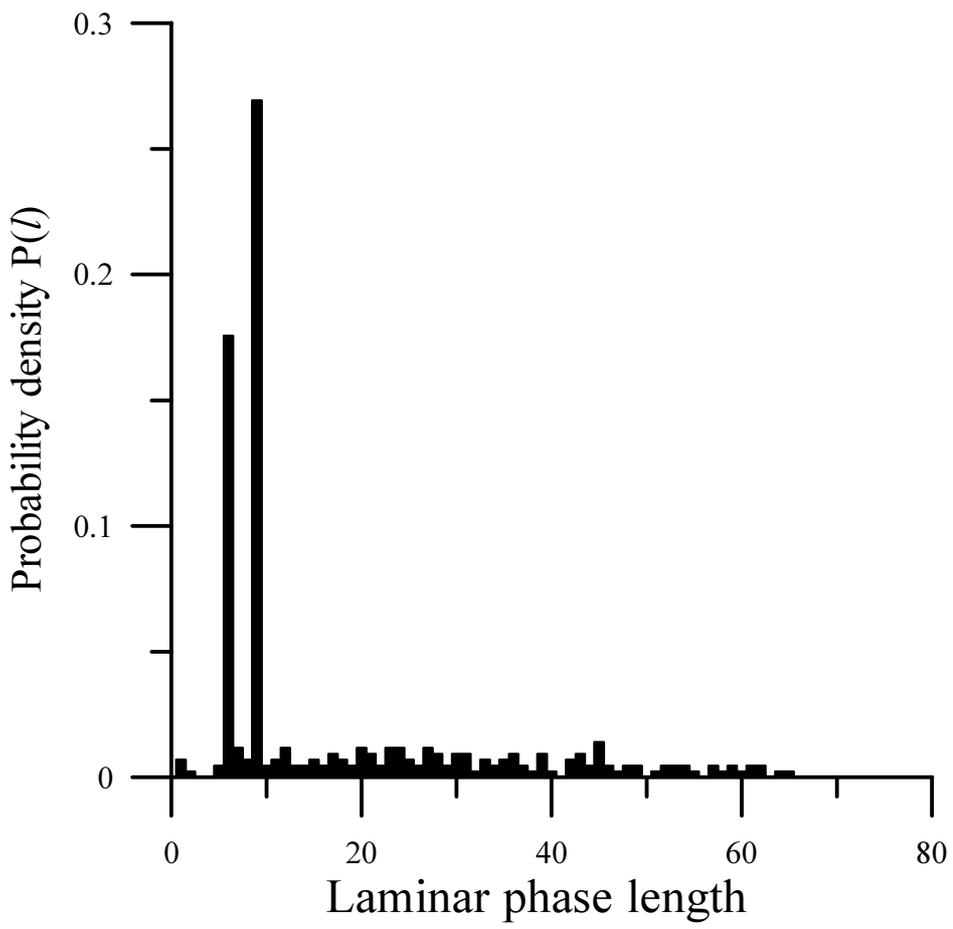
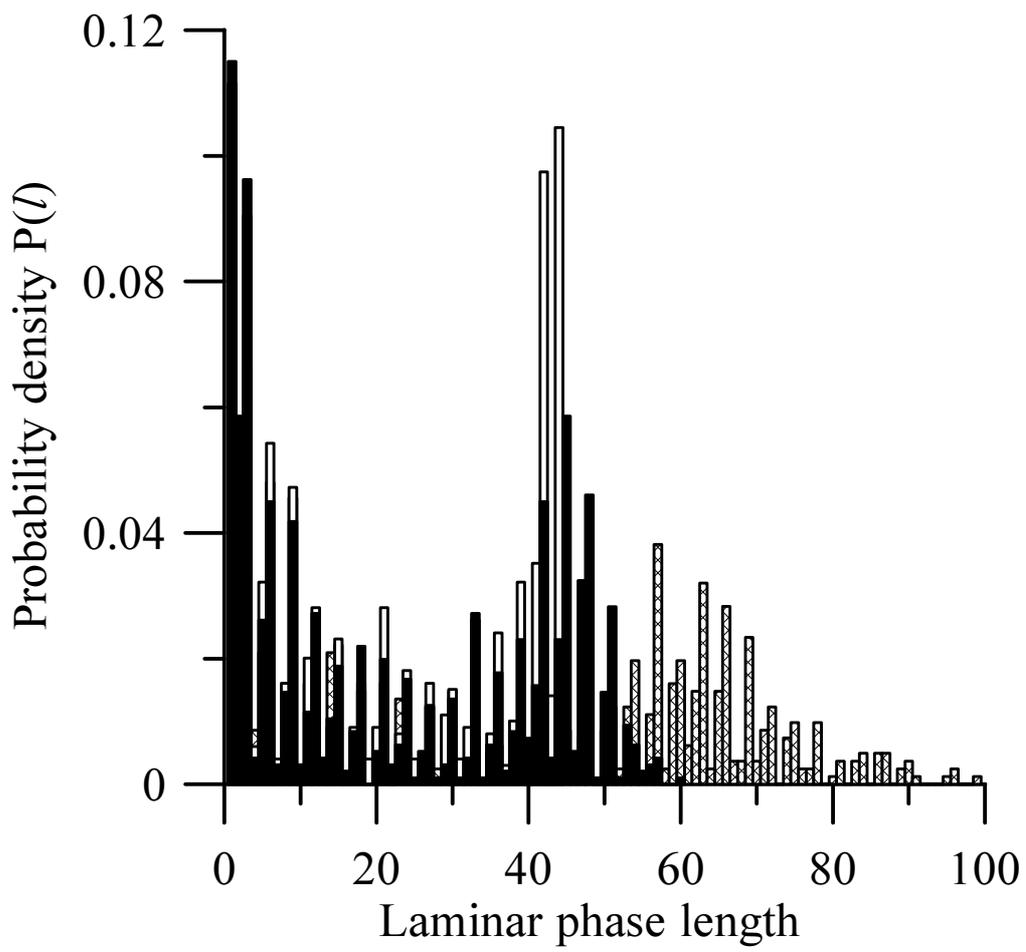
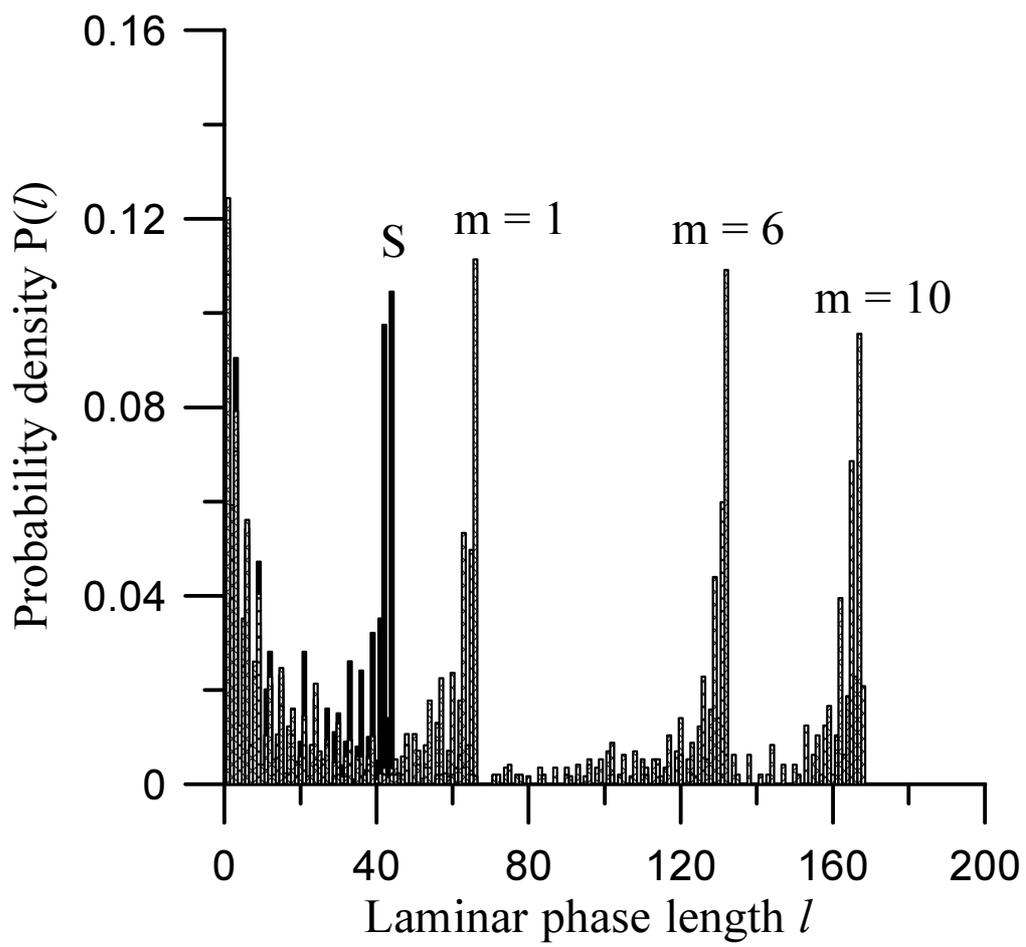
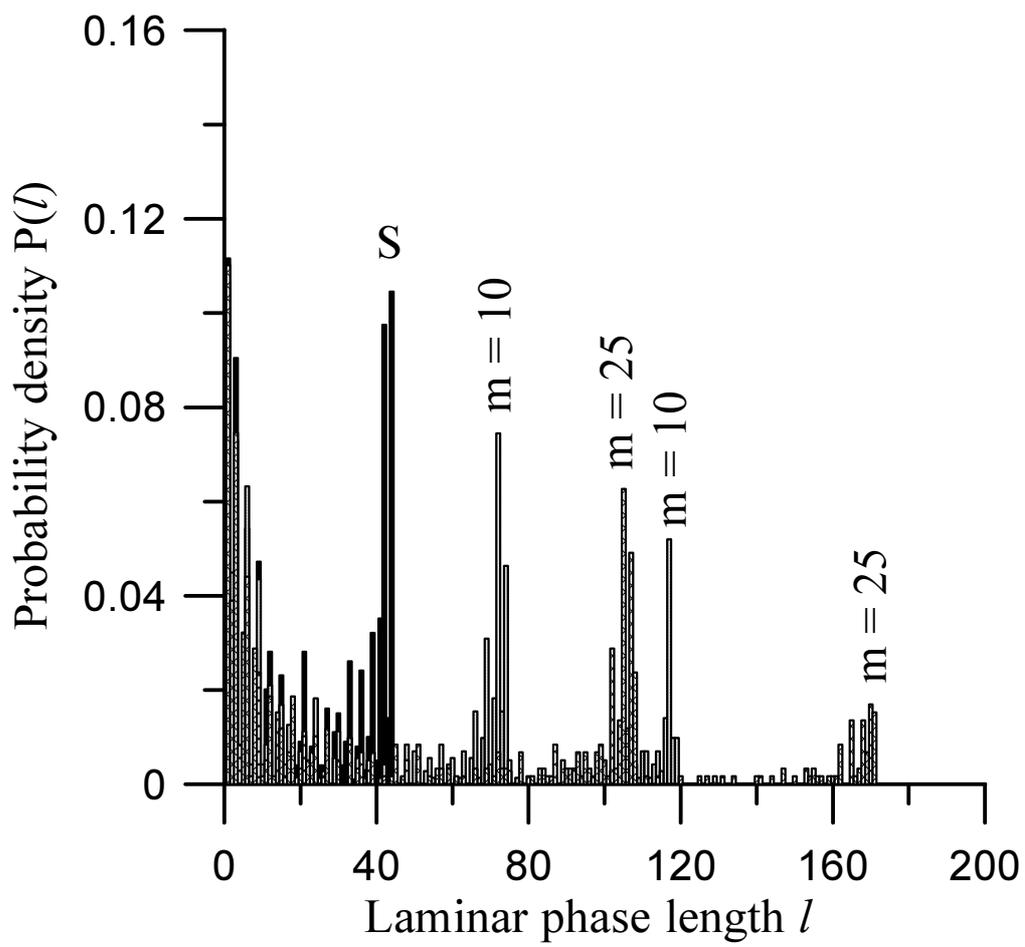
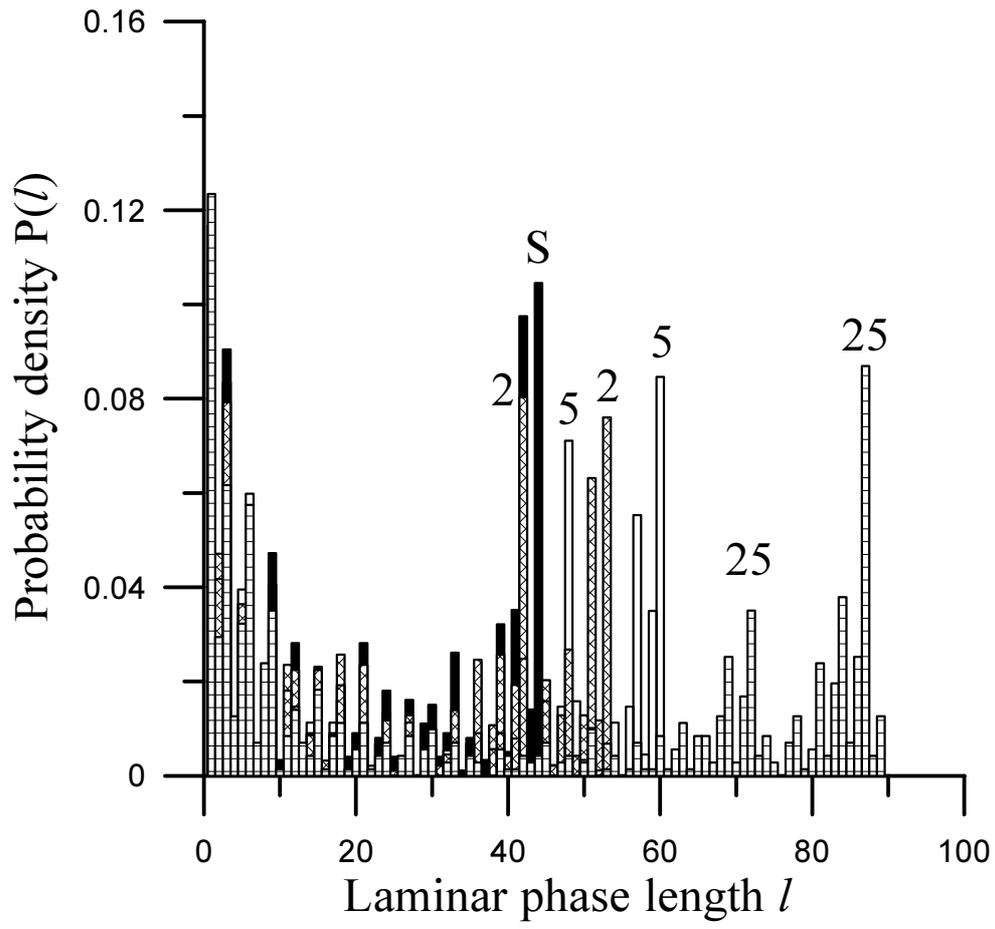
\end{document}